\documentclass[twocolumn,preprintnumbers,prl]{revtex4}
\usepackage{amsfonts}
\usepackage[T1]{fontenc}
\usepackage{amsmath,amsbsy,amssymb,graphicx}
\usepackage{times}

\begin{document}

\preprint{Journal of the Physical Society of Japan (to be published)}
\title{{\Large Quasi-Topological Insulator and Trigonal Warping }\\
{\Large in Gated Bilayer Silicene}}
\author{Motohiko Ezawa}
\affiliation{Department of Applied Physics, University of Tokyo, Hongo 7-3-1, 113-8656,
Japan }

\let\mathbf=\boldsymbol
\begin{abstract}
Bilayer silicene has richer physical properties than bilayer graphene due to
its buckled structure together with its trigonal symmetric structure. 
The trigonal symmetry originates in a particular way of hopping between two silicenes. It
is a topologically trivial insulator since it carries a trivial $\mathbb{Z}_{2}$ topological charge. Nevertheless, its physical properties are more
akin to those of a topological insulator than those of a band insulator.
Indeed, a bilayer silicene nanoribbon has edge modes which are almost
gapless and helical. We may call it a quasi-topological insulator. An
important observation is that the band structure is controllable by applying
the electric field to a bilayer silicene sheet. We investigate the energy
spectrum of bilayer silicene under electric field. Just as monolayer
silicene undergoes a phase transition from a topological insulator to a band
insulator at a certain electric field, bilayer silicene makes a transition
from a quasi-topological insulator to a band insulator beyond a certain
critical field. Bilayer silicene is a metal while monolayer silicene is a
semimetal at the critical field. Furthermore we find that there are several
critical electric fields where the gap closes due to the trigonal warping
effect in bilayer silicene.
\end{abstract}

\maketitle

%\date{}

\section{Introduction}

Silicene is a monolayer of silicon atoms forming a two-dimensional honeycomb
lattice\cite{GLayPRL,Kawai,Takamura}. It has a relatively large spin-orbit
(SO) gap, and its intrinsic property is the buckled structure\cite%
{Shiraishi,Ciraci,LiuPRL} owing to a large ionic radius of silicon. Silicene
has richer physical properties than graphene due to this property. First of
all, it is a topological insulator, characterized by a full insulating gap
in the bulk and helical gapless edges\cite{LiuPRL}. Furthermore, the band
structure is controllable by applying the electric field $E_{z}$ to a
silicene sheet\cite{EzawaNJP}: It has been shown that silicene undergoes a
topological phase transition from a topological insulator to a band
insulator as $|E_{z}| $ increases. It is a semimetal at the critical field $%
E_{\text{cr}}$ due to its linear dispersion relation. It has many attractive
and remarkable properties\cite{EzawaSiQHE,EzawaSiPRL,EzawaSiTube}. Silicene
may be the most promising material now available as a topological insulator%
\cite{Hasan,Qi,KaneMele,Wu}.

In this paper we analyze the band structure of bilayer silicene, which was
manufactured\cite{Feng} experimentally only recently. Though bilayer
silicene has a trivial $\mathbb{Z}_{2}$ topological charge, we show that its
physical properties are more akin to those of a topological insulator than
those of a band insulator. When we switch off the Rashba SO interaction ($%
\lambda _{\text{R}}=0$) and the interlayer SO interaction ($\lambda _{\text{%
inter}}=0$), bilayer silicene shares the same edge-mode properties with a
topological insulator: It has a full insulating gap in the bulk and helical
gapless edges. However, helical gapless edge modes are not topologically
protected because of a possible mixing between the two gapless modes present
in each edge. Indeed, a small gap opens when these interactions are taken
into account. Nevertheless, the physical properties are very much similar to
those of a topological insulator. This is particularly so under electric
field, as we shall soon see. It would be reasonable to call such a system a
quasi-topological insulator.

Here we recall a recent proposal of bilayer graphene\cite{Prada} with rather
unphysical parameters for study of topological insulator. The model
Hamiltonian is very similar to ours, and there are much in common.
Nevertheless, the great merit of our theory is that silicene is a realistic
material, and any experimental test on our results is feasible. Furthermore,
it is remarkable that the band structure is controllable by applying the
electric field to bilayer silicene. Just as monolayer silicene undergoes a
phase transition from a topological insulator to a band insulator at a
certain electric field\cite{EzawaNJP}, bilayer silicene makes a transition
from a quasi-topological insulator to a simple band insulator beyond a
certain critical field. It is a metal at the critical field. There exists
also a new feature. We find that the band gap closes a few times when we
increase electric field, which is absent in the case of monolayer silicene.
It is a trigonal warping effect.

Bilayer silicene has an additional degree of freedom how to stack two
buckled monolayer silicenes. There are four types of stacking even if we
concentrate on the Bernal (AB) stacking (Fig.1). For example, the forward
stacking has an electron-hole symmetry but the backward stacking does not.
It is possible to determine the type of stacking of a sample by experimental
measurement.

This paper is organized as follows. In Section II we introduce the
Hamiltonian for bilayer silicene and study the band structure of a
nanoribbon. We point out by numerical calculation that the edge modes are
almost gapless and helical though bilayer silicene has a trivial $\mathbb{Z}%
_{2}$ topological charge. In Section III we derive the effective Dirac
theory to describe the low-energy physics around the K and K' points. In
Section IV we investigate the band structure under homogeneous electric
field based on the effective Dirac theory. Bilayer silicene is shown to
become metallic at the critical electric field, though the band structure
depends considerably on the type of stacking of two buckled silicenes.

\section{Silicene and Bilayer Silicene}

Silicene consists of a honeycomb lattice of silicon atoms with two
sublattices made of A sites and B sites. Due to the buckled structure the
two sublattice planes are separated by a distance, which we denote by $2\ell 
$ with $\ell =0.23$\AA . The states near the Fermi energy are $\pi $
orbitals residing near the K and K' points at opposite corners of the
hexagonal Brillouin zone. We refer to the K or K' point also as the K$_{\eta
}$ point with the valley index $\eta =\pm 1$.

The monolayer silicene system is described by the four-band
second-nearest-neighbor tight binding model\cite{LiuPRB},

\begin{align}
H_{\text{SL}}& =-t\sum_{\left\langle i,j\right\rangle \alpha }c_{i\alpha
}^{\dagger }c_{j\alpha }+i\frac{\lambda _{\text{SO}}}{3\sqrt{3}}%
\sum_{\left\langle \!\left\langle i,j\right\rangle \!\right\rangle \alpha
\beta }\nu _{ij}c_{i\alpha }^{\dagger }\sigma _{\alpha \beta }^{z}c_{j\beta }
\notag \\
& -i\frac{2}{3}\lambda _{\text{R}}\sum_{\left\langle \!\left\langle
i,j\right\rangle \!\right\rangle \alpha \beta }\mu _{i}c_{i\alpha }^{\dagger
}\left(\mathbf{\sigma }\times \hat{\mathbf{d}}_{ij}\right) _{\alpha \beta
}^{z}c_{j\beta }.  \label{BasicHamil}
\end{align}%
where $c_{i\alpha }^{\dagger }$ creates an electron with spin polarization $%
\alpha $ at site $i$, and $\left\langle i,j\right\rangle /\left\langle
\!\left\langle i,j\right\rangle \!\right\rangle $ run over all the
nearest/second-nearest neighbor hopping sites. The first term represents the
usual nearest-neighbor hopping with the transfer energy $t=1.6$eV. The
second term represents the effective SO coupling with $\lambda _{\text{SO}%
}=3.9$meV, where $\mathbf{\sigma }=(\sigma _{x},\sigma _{y},\sigma _{z})$ is
the Pauli matrix of spin, and $\nu _{ij}=+1$ if the
second-nearest-neighboring hopping is anticlockwise and $\nu _{ij}=-1$ if it
is clockwise with respect to the positive z axis. The third term represents
the Rashba SO coupling with $\lambda _{\text{R}}=0.7$meV, where $\mu_i=\pm 1 
$ for the A (B) site, and $\hat{\mathbf{d}}_{ij}=\mathbf{d}_{ij}/\left\vert 
\mathbf{d}_{ij}\right\vert $ with $\mathbf{d}_{ij}$ the vector connecting
two sites $i$ and $j$ in the same sublattice. Monolayer silicene has been
shown to be a topological insulator\cite{LiuPRL,EzawaNJP}.

\begin{figure}[t]
\centerline{\includegraphics[width=0.4\textwidth]{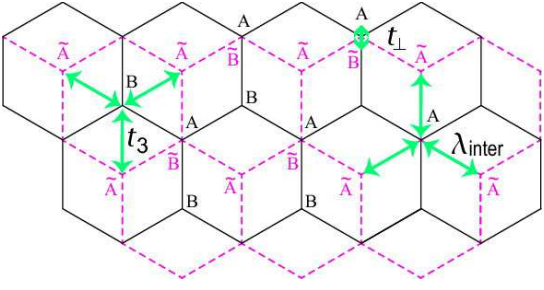}}
\caption{(Color online) Schematic illustration of AB stacking bilayer
honeycomb lattice. Bond connecting sites $A$, $B$ in the top layer is
indicated by solid line, while bond connecting sites $\tilde{A}$, $\tilde{B}$
in the bottom layer by dashed line. The interlalyer interactions are
indicated by arrows together with the couplings $t_{\bot }$, $t_{3}$ and $%
\protect\lambda _{\text{inter}}$.}
\label{FigIllustBilayer}
\end{figure}

We model a bilayer silicene as two coupled buckled hexagonal lattices
including inequivalent sites $A,B$ in the top layer and $\tilde{A},\tilde{B}$
in the bottom layer, respectively. These are arranged according to Bernal ($%
A $-$\tilde{B}$) stacking, as illustrated in Fig.\ref{FigIllustBilayer}.
There are four types of Bernal stacking in bilayer silicene, since the top
and bottom layers can be buckled upward and downward independently.
Irrespective of the type, the bilayer silicene system is described by the
eight-band second-nearest-neighbor tight binding model,%
\begin{equation}
H_{\text{BL}}=H_{\text{SL}}^{\text{T}}+H_{\text{SL}}^{\text{B}}+H_{\text{%
inter}}  \label{BLG}
\end{equation}%
\begin{eqnarray}
H_{\text{inter}} &=&t_{\perp }\sum_{i\in A,j\in \tilde{B}}\left( c_{i\alpha
}^{\dagger }c_{j\alpha }+\text{h.c.}\right) +t_{3}\sum_{i\in B,j\in \tilde{A}%
}\left( c_{i\alpha }^{\dagger }c_{j\alpha }+\text{h.c.}\right)  \notag \\
&&+i\lambda _{\text{inter}}\sum_{i\in A,j\in \tilde{A}}c_{i\alpha }^{\dagger
}\left( \mathbf{\sigma }\times \hat{\mathbf{d}}_{ij}\right) _{\alpha \beta
}^{z}c_{j\beta },
\end{eqnarray}%
where $H_{\text{SL}}^{\text{T(B)}}$ is the Hamiltonian (\ref{BasicHamil})
for the top (T) or bottom (B) layer; $H_{\text{inter}}$ is the interlayer
Hamiltonian, where the first term is the nearest-neighbor interlayer
vertical hopping between $A$ sites of the top layer and $\tilde{B}$ sites of
the bottom layer with the coupling $t_{\perp }$, and the second term is the
next-nearest-neighbor interlayer hopping between $B$ sites of the top layer
and $\tilde{A}$ sites of the bottom layer with the coupling $t_{3}$. There
are three $t_{3}$ hopping vectors from one site, as implies the trigonal
symmetry of the bilayer system. The parameters $t_{\perp }$ and $t_{3}$
depend on the type of Bernal stacking, but we expect $t_{\perp }\gtrsim
t_{3}\sim 0.1t$ in all of them. The third term describes the interlayer SO
interaction\cite{Dressel,Guinea} with the coupling $\lambda _{\text{inter}}$%
. Its magnitude would be the same order as that in graphene ($\lambda _{%
\text{inter}}\sim 0.5$meV).

\begin{figure}[t]
\centerline{\includegraphics[width=0.4\textwidth]{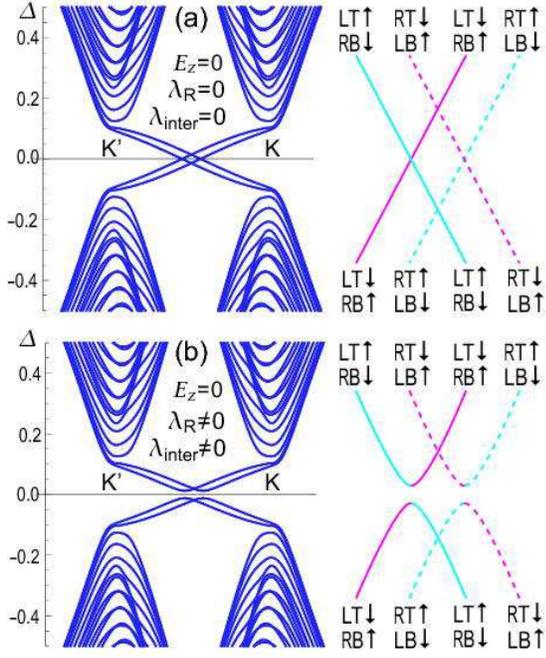}}
\caption{(Color online) One-dimensional band structure for a bilayer
silicene nanoribbon. It consists of the bulk modes and the edge modes. The
vertical axis is the energy gap $\Delta $ in unit of $t$. The horizontal
axis is $k_{x}$. (a) When the Rashba and interlayer interactions are
switched off ($\protect\lambda _{\text{R}}=\protect\lambda _{\text{inter}}=0$%
), there are eight bands crossing the gap, which are edge modes. There are
four due to the left (L) and right (R) edges of the top (T) and bottom (B)
silicenes, with each of them doubly degenerated with respect to the spin $%
\uparrow $ and $\downarrow $. The gapless edge modes are helical. (b) When
these interactions are switched on ($\protect\lambda _{\text{R}}\protect%
\lambda _{\text{inter}}\neq 0$), a crossing turns out to be an anticrossing
due to a spin mixing, as induces a small gap. The edge modes are almost
helical except for the anticrossing point. In fact, spin mixing occurs only
in the vicinity of the anticrossing point. We have taken $\protect\lambda _{%
\text{R}}/\protect\lambda _{\text{SO}}=1/2$ and $\protect\lambda _{\text{%
inter}}/\protect\lambda _{\text{SO}}=1/4$ for illustration to emphasize the
gap.}
\label{FigBiRibbonBand}
\end{figure}

It is readily shown that the band structure of the Hamiltonian (\ref{BLG})
is gapped, and hence bilayer silicene is an insulator. To study what type of
insulator it is, we have analyzed the energy bands of a bilayer silicene
nanoribbon based on the tight-binding model (\ref{BLG}), whose result we
give in Fig.\ref{FigBiRibbonBand}. The band structure consists of the bulk
modes and the edge modes, where the bulk modes are gapped. There exists an
intriguing feature with respect to the edge modes.

When we switch off the Rashba and interlayer interactions ($\lambda _{\text{R%
}}=\lambda _{\text{inter}}=0$) there emerge eight helical zero-energy modes
in the edges [Fig.\ref{FigBiRibbonBand}(a)]. This is what we expect since
the number is simply twice as much as that in monolayer silicene\cite%
{LiuPRL,EzawaNJP}. Namely, four helical zero-energy modes arise from each
monolayer silicene. Indeed, according to the bulk-boundary correspondence
rule, and since the Chern number of the bilayer system is shown to be twice
that of the monolayer, the number of helical edge states should also be
double.

However, when we switch on these interactions ($\lambda _{\text{R}}\lambda _{%
\text{inter}}\neq 0$), a spin mixing occurs, as turns a two-band crossing
into a two-band anticrossing and opens a small gap for the edge modes [Fig.%
\ref{FigBiRibbonBand}(b)]. This makes a sharp contrast to the monolayer
case, where the gapless edge modes are protected topologically against
perturbations. Despite each layer behaves as a topological insulator, an
even number of such layers renders the overall system topologically trivial
that is characterized by a vanishing $\mathbb{Z}_{2}$ charge. The
topological triviality of a bilayer system may be understood as follows. In
the bilayer system, backscattering between channels at the same edge moving
in opposite directions with opposite spins is not forbidden, in contrast to
the case for the monolayer system. We note that a similar behavior has been
pointed out in bilayer graphene\cite{Prada}.

Nevertheless, the properties of bilayer silicene are more akin to those of a
topological insulator than those of a band insulator. When $\lambda _{\text{R%
}}=\lambda _{\text{inter}}=0$ the edge modes are purely helical, which is
one of the characteristic feature of a topological insulator. When $\lambda
_{\text{R}}\lambda _{\text{inter}}\neq 0$ a spin mixing occurs only around
the anticrossing point, and the edge modes are almost helical away from the
point, as we can confirm numerically. As we shall soon show, by applying a
critical electric field, the gap becomes precisely zero at the K and K'
points: Bilayer silicene becomes a metal due to a parabolic dispersion, just
as monolayer silicene becomes a semimetal\cite{EzawaNJP} due to a linear
dispersion. It is reasonable to call such an object a quasi-topological
insulator.

\section{Low-Energy Dirac Theory}

\begin{figure}[t]
\centerline{\includegraphics[width=0.3\textwidth]{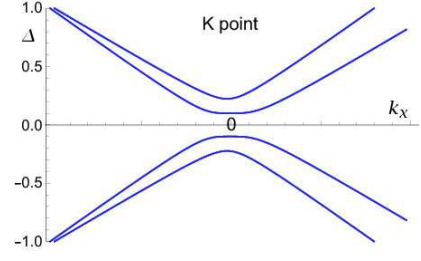}}
\caption{(Color online) The band structure of bilayer silicene near the K
point based on the Dirac Hamiltonian (\protect\ref{DiracHamil}). The
vertical axis is the energy gap $\Delta $ in unit of $t$. The horizontal
axis is $k_{x}$. It is clearly gapped. There is no symmetry for $%
k_{x}\longleftrightarrow -k_{x}$ due to the $t_{3}$ hopping term, reflecting
the trigonal warping. }
\label{FigE0Band}
\end{figure}

We analyze the physics of electrons near the Fermi energy based on the
low-energy effective Hamiltonian derived from the tight binding model (\ref%
{BLG}). It is described by the Dirac theory around the $K_{\eta }$ point as%
\begin{equation}
H_{\eta }=\hbar v_{\text{F}}(k_{x}\tau _{x}^{\text{AB}}-\eta k_{y}\tau _{y}^{%
\text{AB}})+H_{\text{trans}}^{\text{inter}}+H_{\text{SO}}^{\text{intra}}+H_{%
\text{SO}}^{\text{inter}},  \label{DiracHamil}
\end{equation}%
with%
\begin{align}
H_{\text{trans}}^{\text{inter}}& =\frac{t_{\perp }}{2}(\tau _{x}^{\text{AB}%
}\tau _{x}^{\text{layer}}-\tau _{y}^{\text{AB}}\tau _{y}^{\text{layer}}) 
\notag \\
& \qquad +\frac{t_{3}}{2}[k_{x}(\tau _{x}^{\text{AB}}\tau _{x}^{\text{layer}%
}+\tau _{y}^{\text{AB}}\tau _{y}^{\text{layer}}),
\end{align}%
and%
\begin{equation}
H_{\text{SO}}^{\text{intra}}=-\eta \tau _{z}^{\text{AB}}[\lambda _{\text{SO}%
}\sigma _{z}+a\lambda _{\text{R}}\left( k_{y}\sigma _{x}-k_{x}\sigma
_{y}\right) ],  \label{HamilH11}
\end{equation}%
and 
\begin{equation}
H_{\text{SO}}^{\text{inter}}=\lambda _{4}^{\text{inter}}\tau _{z}^{\text{AB}%
}(\tau _{y}^{\text{layer}}\sigma _{x}+\eta \tau _{x}^{\text{layer}}\sigma
_{y}),
\end{equation}%
where $\tau _{a}^{\text{AB}}$ ($\tau _{a}^{\text{layer}}$) is the Pauli
matrix of the sublattice (layer) pseudospin, $v_{\text{F}}=\frac{\sqrt{3}}{2}%
at=5.5\times 10^{5}$m/s is the Fermi velocity, and $a=3.86$\AA\ is the
lattice constant. We show the band structure in Fig.\ref{FigE0Band}.

The Hamiltonian contains many parameters. Actually, $\lambda _{\text{R}}$
and $\lambda _{\text{inter}}$ are small constants with respect to the other
parameters. Hence, it is a good approximation to set $\lambda _{\text{R}%
}=\lambda _{\text{inter}}=0$ in most cases. In what follow, although we
carry out numerical analysis by including the nonzero effects of $\lambda _{%
\text{R}}$ and $\lambda _{\text{inter}}$, we develop an analytic formalism
by assuming $\lambda _{\text{R}}=\lambda _{\text{inter}}=0$. It enables us
to make a simple and clear physical picture on the basis of analytic
formulas. For instance, the spin $s_{z}=\pm 1$ becomes a good quantum number
in this simplification, where the spin-chirality is given by $s=\eta s_{z}$.
We can always check the validity of the approximation numerically.

We rewrite down the Hamiltonian $H_{+}$ explicitly\ in the basis $\left\{
\psi _{A\uparrow },\psi _{\tilde{B}\uparrow },\psi _{\tilde{A}\uparrow
},\psi _{B\uparrow }\right\} ^{t}$, 
\begin{equation}
H_{+}=\eta \left( 
\begin{array}{cc}
H_{11} & H_{12} \\ 
H_{21} & H_{22}%
\end{array}%
\right) ,  \label{HamilExpli}
\end{equation}%
where the diagonal blocks are 
\begin{equation}
H_{11}=\left( 
\begin{array}{cc}
\eta E(s,1,1) & \hbar v_{3}k_{+} \\ 
\hbar v_{3}k_{-} & \eta E(s,-1,-1)%
\end{array}%
\right) ,  \label{Hamil11}
\end{equation}%
\begin{equation}
H_{22}=\left( 
\begin{array}{cc}
\eta E(s,1,-1) & \eta t_{\perp } \\ 
\eta t_{\perp } & \eta E(s,-1,1)%
\end{array}%
\right) ,  \label{Hamil22}
\end{equation}%
while the off-diagonal blocks are 
\begin{equation}
H_{12}=H_{21}=\left( 
\begin{array}{cc}
0 & \hbar v_{\text{F}}k_{-} \\ 
\hbar v_{\text{F}}k_{+} & 0%
\end{array}%
\right) ,
\end{equation}%
with $v_{3}=\frac{\sqrt{3}}{2}at_{3}$ and $k_{\pm }=k_{x}\pm ik_{y}$. The
diagonal elements are%
\begin{equation}
E(s,t_{z}^{\text{AB}},t_{z}^{\text{layer}})=-t_{z}^{\text{AB}}s\lambda _{%
\text{SO}},  \label{DiagoE}
\end{equation}%
where $t_{z}^{\text{layer}}=\pm 1$ for the upper (lower) silicene sheet. We
have introduced this parameter for a later convenience, though it is
irrelevant here.

We make a further simplification of the theory by reducing the $4\times 4$
Hamiltonian to the $2\times 2$ Hamiltonian, which is valid as the effective
two-band Hamiltonian for the lower energy bands\cite{LLBilayer}. Provided $%
t_{3}<t_{\perp }$, the reduced Hamiltonian is given by%
\begin{equation}
H_{\text{eff}}=H_{11}-H_{12}G_{22}^{0}H_{21},  \label{perturb}
\end{equation}%
where $G_{22}^{0}$ is the $2\times 2$ Green function, 
\begin{equation}
G_{\alpha \alpha }^{0}=(H_{\alpha \alpha }-\varepsilon )^{-1},
\end{equation}%
with the energy $\varepsilon $ from the Fermi level. The effective
Hamiltonian around the K$_{\eta }$ point reads%
\begin{align}
H_{\text{eff}}^{\eta }=& h_{B}+h_{w}-s\lambda _{\text{SO}}\left( 
\begin{array}{cc}
1 & 0 \\ 
0 & -1%
\end{array}%
\right)  \notag \\
& -\frac{\hbar ^{2}v_{\text{F}}^{2}}{t_{\perp }^{2}}s\lambda _{\text{SO}%
}\left( 
\begin{array}{cc}
k_{-}k_{+} & 0 \\ 
0 & -k_{+}k_{-}%
\end{array}%
\right) ,
\end{align}%
with%
\begin{align}
h_{B}=& -\frac{\hbar ^{2}v_{\text{F}}^{2}}{t_{\perp }}\left( 
\begin{array}{cc}
0 & k_{-}^{2} \\ 
k_{+}^{2} & 0%
\end{array}%
\right) , \\
h_{w}=& \eta \hbar v_{3}\left( 
\begin{array}{cc}
0 & k_{+} \\ 
k_{-} & 0%
\end{array}%
\right) ,
\end{align}%
where we have used $\left\vert \varepsilon \right\vert \ll t_{\perp }$. The
Hamiltonian $h_{B}$ describes the quadratic Dirac Hamiltonian, which is
familiar in bilayer graphene. The Hamiltonian $h_{w}$ describes the trigonal
warping effects. We note that the reduction into the $2\times 2$ Hamiltonian
is not available when $t_{3}>t_{\perp }$.

\section{Bilayer Silicene in Electric Field}

\label{SecElectField}

We take a bilayer silicene sheet on the $xy$-plane, and apply the electric
field $E_{z}$ perpendicular to the plane. It generates a staggered
sublattice potential $\varpropto \pm 2\ell E_{z}$ between silicon atoms at A
sites and B sites in a single silicene sheet, and also a staggered
sublattice potential $\varpropto \pm 2LE_{z}$ between two silicene sheets.
How they enter into the Hamiltonian depends on the type of four Bernal
stacking. They are

\begin{figure}[t]
\centerline{\includegraphics[width=0.5\textwidth]{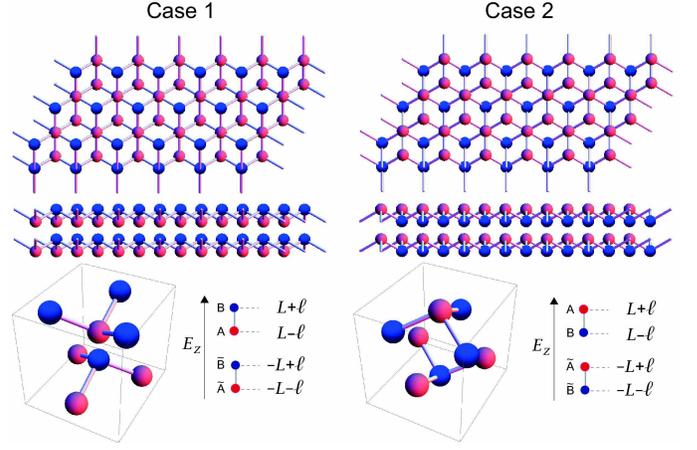}}
\caption{(Color online) Illustration of the forward Bernal (AB) stacking
bilayer silicene. The order of sites is $B,A,\tilde{B},\tilde{A}$ for the
case 1 and $A,B,\tilde{A},\tilde{B}$ for the case 2 from the top to the
bottom.}
\label{FigFAB}
\end{figure}

1) The top layer buckles downwardly and the bottom layer buckles downwardly
(Fig.\ref{FigFAB}).

2) The top layer buckles upwardly and the bottom layer buckles upwardly (Fig.%
\ref{FigFAB}).

\begin{figure}[t]
\centerline{\includegraphics[width=0.5\textwidth]{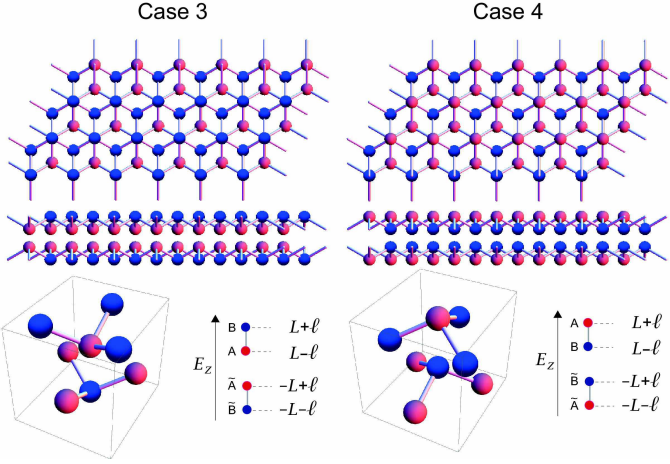}}
\caption{(Color online) Illustration of the backward Bernal (AB) stacking
bilayer silicene. The order of sites is $B,A,\tilde{A},\tilde{B}$ for the
case 3 and $A,B,\tilde{B},\tilde{A}$ for the case 4 from the top to the
bottom.}
\label{FigBAB}
\end{figure}

3) The top layer buckles upwardly and the bottom layer buckles downwardly
(Fig.\ref{FigBAB}).

4) The top layer buckles downwardly and the bottom layer buckles upwardly
(Fig.\ref{FigBAB}).

\noindent%
Here, upward (downward) means that A site is higher (lower) than B
on the same layer.

Silicon atoms prefer having the neighboring atoms in regular tetrahedron
directions, and hence the case 1 seems most likely. The cases 3 and 4 are
identical by reversing the whole system, and the energy spectrum is
identical between them. Thus it is enough to investigate the cases 1, 2 and
3. These three systems can not be transformed only by the translation. It is
necessary to rotate 120 degrees after translation. We call the case 1 and 2
(3 and 4) as the forward (backward) Bernal stacking. The forward (backward)
Bernal stacking bilayer silicene is shown in Fig.\ref{FigFAB} (\ref{FigBAB}%
), where the order of sites is $A,B,\tilde{A},\tilde{B}$ ($B,A,\tilde{A},%
\tilde{B}$). The electric energy depends on the order. The magnitude of the
interlayer SO interaction $\lambda _{\text{inter}}$ depends on the stacking
type.

The effective Hamiltonian for the forward stacking is given by%
\begin{equation}
H=H_{\text{BL}}+(L\tau _{z}^{\text{layer}}+\ell \tau _{z}^{\text{AB}%
})\sum_{i\alpha }\eta _{i}E_{z}^{i}c_{i\alpha }^{\dagger }c_{i\alpha },
\label{HamilElectFor}
\end{equation}%
while the one for the backward stacking is given by%
\begin{equation}
H=H_{\text{BL}}+(L-\ell \tau _{z}^{\text{AB}})\tau _{z}^{\text{layer}%
}\sum_{i\alpha }\eta _{i}E_{z}^{i}c_{i\alpha }^{\dagger }c_{i\alpha },
\label{HamilElectBack}
\end{equation}%
where $H_{\text{BL}}$ is given by (\ref{BLG}). Accordingly, the Dirac
Hamiltonian (\ref{DiracHamil}) is modified as%
\begin{equation}
H_{\eta }=H_{\eta }^{0}+E(s,t_{z}^{\text{AB}},t_{z}^{\text{layer}}),
\label{DiracHamilE}
\end{equation}%
where $H_{\eta }^{0}$ is the Dirac Hamiltonian (\ref{DiracHamil}), and 
\begin{equation}
E(s,t_{z}^{\text{AB}},t_{z}^{\text{layer}})=-s\lambda _{\text{SO}}t_{z}^{%
\text{AB}}+(L\tau _{z}^{\text{layer}}+\ell \tau _{z}^{\text{AB}})E_{z}
\label{DiagEfor}
\end{equation}%
for the forward stacking, and%
\begin{equation}
E(s,t_{z}^{\text{AB}},t_{z}^{\text{layer}})=-s\lambda _{\text{SO}}t_{z}^{%
\text{AB}}+\tau _{z}^{\text{layer}}(L-\ell \tau _{z}^{\text{AB}})E_{z}
\label{DiagoEback}
\end{equation}%
for the backward stacking. The modification is very simple for the four-band
theory: The electric field is incorporated into the Dirac Hamiltonian (\ref%
{HamilExpli}) only by changing the diagonal elements $E(s,t_{z}^{\text{AB}%
},t_{z}^{\text{layer}})$ in (\ref{Hamil11}) and (\ref{Hamil22}) with (\ref%
{DiagEfor}) or (\ref{DiagoEback}).

We first investigate the case 1, and then the cases 3 and 4. The case 2 is
investigated at the end since the two-band Hamiltonian is unavailable.

\subsection{Forward Bernal Stacking in Electric Field: Case 1, $BA$-$\tilde{B%
}\tilde{A}$ type}

\begin{figure}[t]
\centerline{\includegraphics[width=0.4\textwidth]{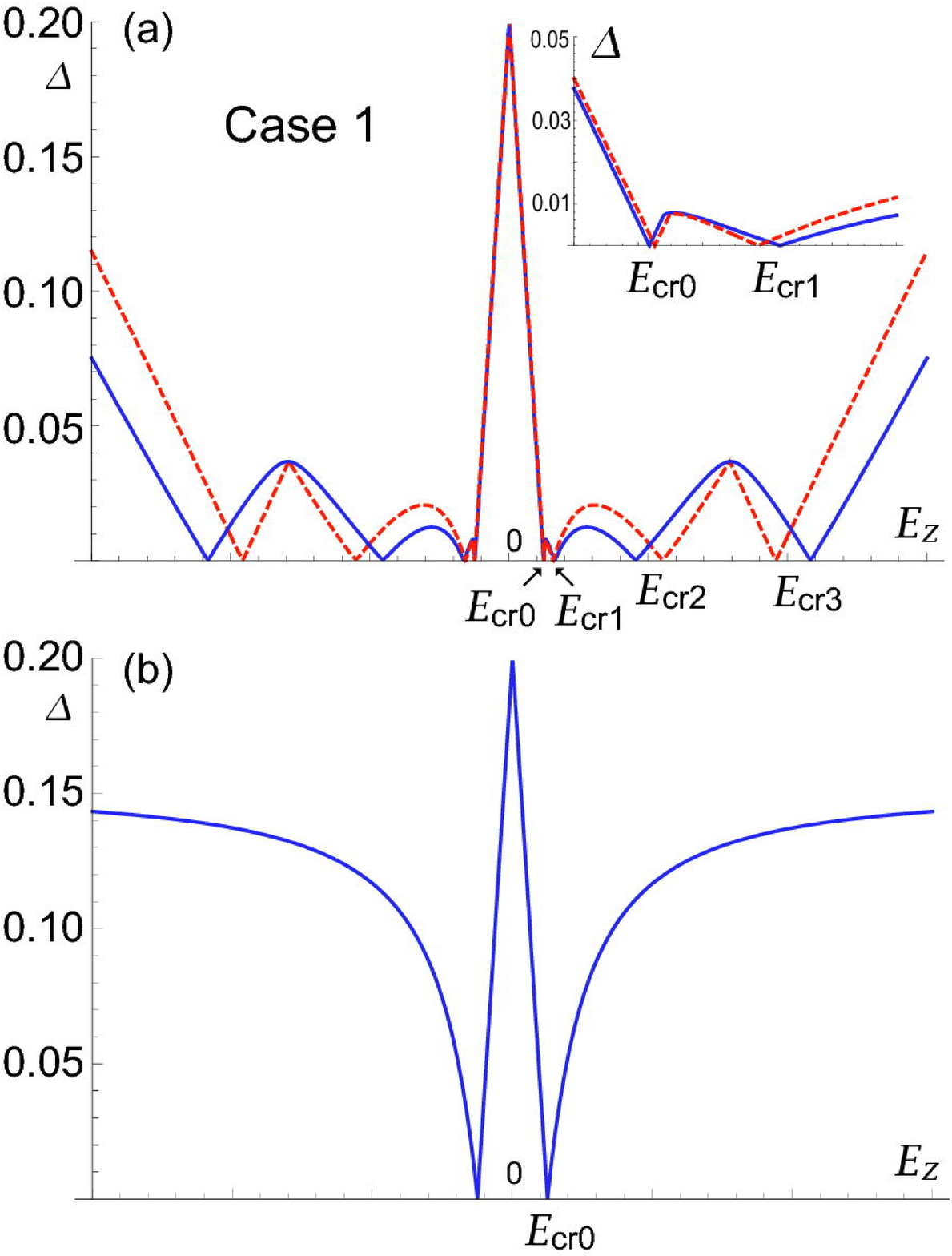}}
\caption{(Color online) (a) Band gap $\Delta $ of the forward stacking
silicene as a function of the electric field $E_{z}$ for the case 1. The gap
closes at the critical points $E_{\text{cr}0},E_{\text{cr}1},E_{\text{cr}%
2},E_{\text{cr}3}$. Bilayer silicene is an insulator except for these
points. We note that the critical points $E_{\text{cr}2}$ and $E_{\text{cr}%
3} $ are missed in the effective two-band theory. (b) The band gap in the
limit $t_{3}\rightarrow 0$, where all the critical points collap to one.
Solid (dashed) curve is the band gap obtained based on the Dirac Hamiltonian
(\protect\ref{DiracHamilE}) by taking $\protect\lambda _{\text{R}}/\protect%
\lambda _{\text{SO}}=1/2$, $\protect\lambda _{\text{inter}}/\protect\lambda %
_{\text{SO}}=1/4$ and $\protect\lambda _{\text{SO}}/t=1/10$ for illustration
to emphasize the effects ($\protect\lambda =\protect\lambda _{\text{inter}%
}=0 $ for comparison). They show qualitatively the same behavior.}
\label{FigEGap}
\end{figure}

\begin{figure}[t]
\centerline{\includegraphics[width=0.3\textwidth]{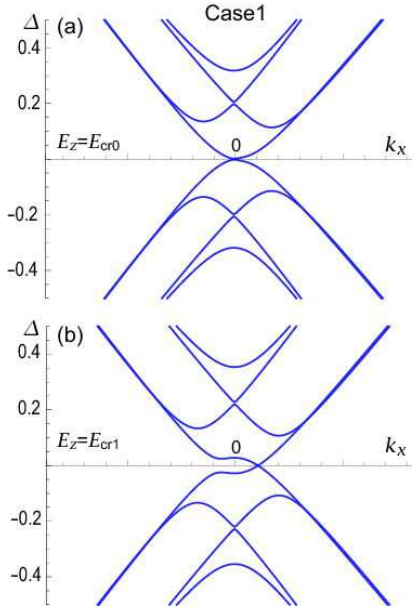}}
\caption{(Color online) Band structure of the forward Bernal (AB) stacking
bilayer silicene (a) at the critical field $E_{\text{cr}0}$ and (b) at the
critical field $E_{\text{cr}1}$ for the case 1. The horizontal axis is $%
k_{x} $. (a) The gapless mode is parabolic, showing that it is a metal at $%
E=E_{\text{cr}0}$. (b) At $E=E_{\text{cr}1}$, there is no symmetry for $%
k_{x}\longleftrightarrow -k_{x}$ due to the $t_{3}$ hopping term, reflecting
the trigonal warping. A 3-dimensional figure is given in Fig.\protect\ref%
{FigFEc3D}. }
\label{FigFEc}
\end{figure}

In the case 1 of the forward Bernal stacking, the effective two-band
Hamiltonian under homogeneous electric field is derived as 
\begin{align}
H_{\text{eff}}^{\eta }=& h_{B}+h_{w}+m_{\text{D}}\left( 
\begin{array}{cc}
1 & 0 \\ 
0 & -1%
\end{array}%
\right)   \notag \\
& -\frac{\hbar ^{2}v_{\text{F}}^{2}}{t_{\perp }^{2}}\left\{ \left( L+\ell
\right) E_{z}+s\lambda _{\text{SO}}\right\} \left( 
\begin{array}{cc}
k_{-}k_{+} & 0 \\ 
0 & -k_{+}k_{-}%
\end{array}%
\right) ,  \label{HamilForward}
\end{align}%
with the Dirac mass%
\begin{equation}
m_{\text{D}}=\left( L-\ell \right) E_{z}-s\lambda _{\text{SO}}.
\end{equation}%
The energy spectrum has the electron-hole symmetry. The spectrum becomes
gapless when $m_{\text{D}}=0$, or $E_{z}=E_{\text{cr}0}$ with%
\begin{equation}
E_{\text{cr0}}=\frac{s\lambda _{\text{SO}}}{L-\ell },  \label{CritiE0}
\end{equation}%
which occurs at $k_{\pm }=0$. We can show that the band gap is given as 
\begin{equation}
\Delta =2|m_{\text{D}}|,
\end{equation}%
for $|E|<E_{\text{cr0}}$. See also Fig.\ref{FigEGap} which we have derived
by numerical calculations, where we compare the two results with $\lambda _{%
\text{R}}\lambda _{\text{inter}}\neq 0$ and with $\lambda _{\text{R}%
}=\lambda _{\text{inter}}=0$. We show the band structure at the critical
electric field in Fig.\ref{FigFEc}.(a). The band structure at the Fermi energy
is parabolic, which is reminiscence of bilayer graphene. We conclude that
the system is a quasi-topological (band) insulator when the Dirac mass is
negative (positive).

The gap closes at $E_{z}=\pm E_{\text{cr}0}$, where it is a metal due to
gapless modes exhibiting a parabolic dispersion relation: See the band
structure at $E_{z}=E_{\text{cr}0}$ in Fig.\ref{FigFEc}(a). It follows that
up-spin ($s_{z}=+1$) electrons are gapless at the K point ($\eta =+1$),
while down-spin ($s_{z}=-1$) electrons are gapless at the K' point ($\eta
=-1 $). Namely, spins are perfectly up (down) polarized at the K (K') point
under the uniform electric field $E_{z}=E_{\text{cr}0}$. Recall that
monolayer silicene is a semimetal due to a linear dispersion at the critical
point\cite{EzawaNJP}.

\begin{figure}[t]
\centerline{\includegraphics[width=0.3\textwidth]{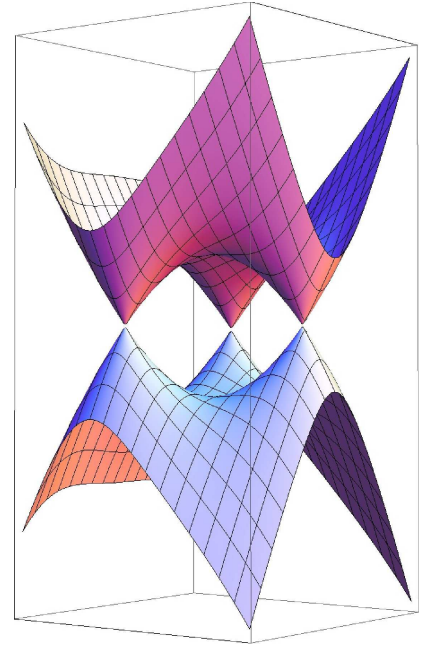}}
\caption{(Color online) Band structure of the forward Bernal (AB) stacking
bilayer silicene at the critical field $E_{\text{cr}1}$ for the case 1. The
Dirac cone is split into three cones, where the band touches the Fermi
level. We have taken $\protect\lambda _{\text{R}}/\protect\lambda _{\text{SO}%
}=1/2$, $\protect\lambda _{\text{inter}}/\protect\lambda _{\text{SO}}=1/4$
and $\protect\lambda _{\text{SO}}/t=1/10$ for illustration.}
\label{FigFEc3D}
\end{figure}

There exists one more critical point implied by the $2\times 2$ Hamiltonian,
where the spectrum become gapless. The condition is that the Hamiltonian (%
\ref{HamilForward}) itself becomes zero, as implies 
\begin{align}
& (L-\ell )E_{z}-s\lambda _{\text{SO}}-\frac{\hbar ^{2}v_{\text{F}}^{2}}{%
t_{\perp }^{2}}\left\{ \left( L+\ell \right) E_{z}+s\lambda _{\text{SO}%
}\right\} k_{-}k_{+}=0, \\
& \hbar v_{3}k_{+}-\frac{\hbar ^{2}v_{\text{F}}^{2}}{t_{\perp }}k_{-}^{2}=0.
\end{align}%
There are two solutions with $k_{y}=0$. The first one is $k_{x}=0$, where $%
E_{z}$ is given by (\ref{CritiE0}). The new solution is given by solving%
\begin{equation}
k_{x}=\pm \frac{t_{\perp }\sqrt{-(L-\ell )E_{z}+s\lambda _{\text{SO}}}}{%
\hbar v_{\text{F}}\sqrt{-(L+\ell )E_{z}-s\lambda _{\text{SO}}}}=\frac{%
v_{3}t_{\perp }}{\hbar v_{\text{F}}^{2}},
\end{equation}%
which yields 
\begin{equation}
E_{\text{cr}1}=\frac{(1+v_{3}^{2}/v_{\text{F}}^{2})s\lambda _{\text{SO}}}{%
(L-\ell )-v_{3}^{2}/v_{\text{F}}^{2}(L+\ell )},
\end{equation}%
as illustrated in Fig.\ref{FigFEc}(b). There are two more solutions by
performing $\pm 2\pi /3$ rotations in the ($k_{x}$,$k_{y}$) plane around ($%
0,0$) with the same critical field $E_{\text{cr}1}$ due to the trigonal
symmetry, as illustrated in Fig.\ref{FigFEc3D}.

Actually there is two more critical point, $E_{\text{cr}2}$ and $E_{\text{cr}%
3}$, which are present in the original Hamiltonian (\ref{HamilExpli}) with (%
\ref{DiagEfor}) but missed in the reduced two-band model (\ref{HamilForward}%
), as we demonstrate in Fig.\ref{FigEGap}(a). Note that $E_{\text{cr}%
i}\rightarrow E_{\text{cr}0}$ for $i=1,2,3$ as $t_{3}\rightarrow 0$. Namely,
these three critical points are generated by the $t_{3}$ hopping effect.

\subsection{Backward Bernal Stacking in Electric Field: Cases 3 and 4, $BA$-$%
\tilde{A}\tilde{B}$ and $AB$-$\tilde{B}\tilde{A}$\ types}

\begin{figure}[t]
\centerline{\includegraphics[width=0.4\textwidth]{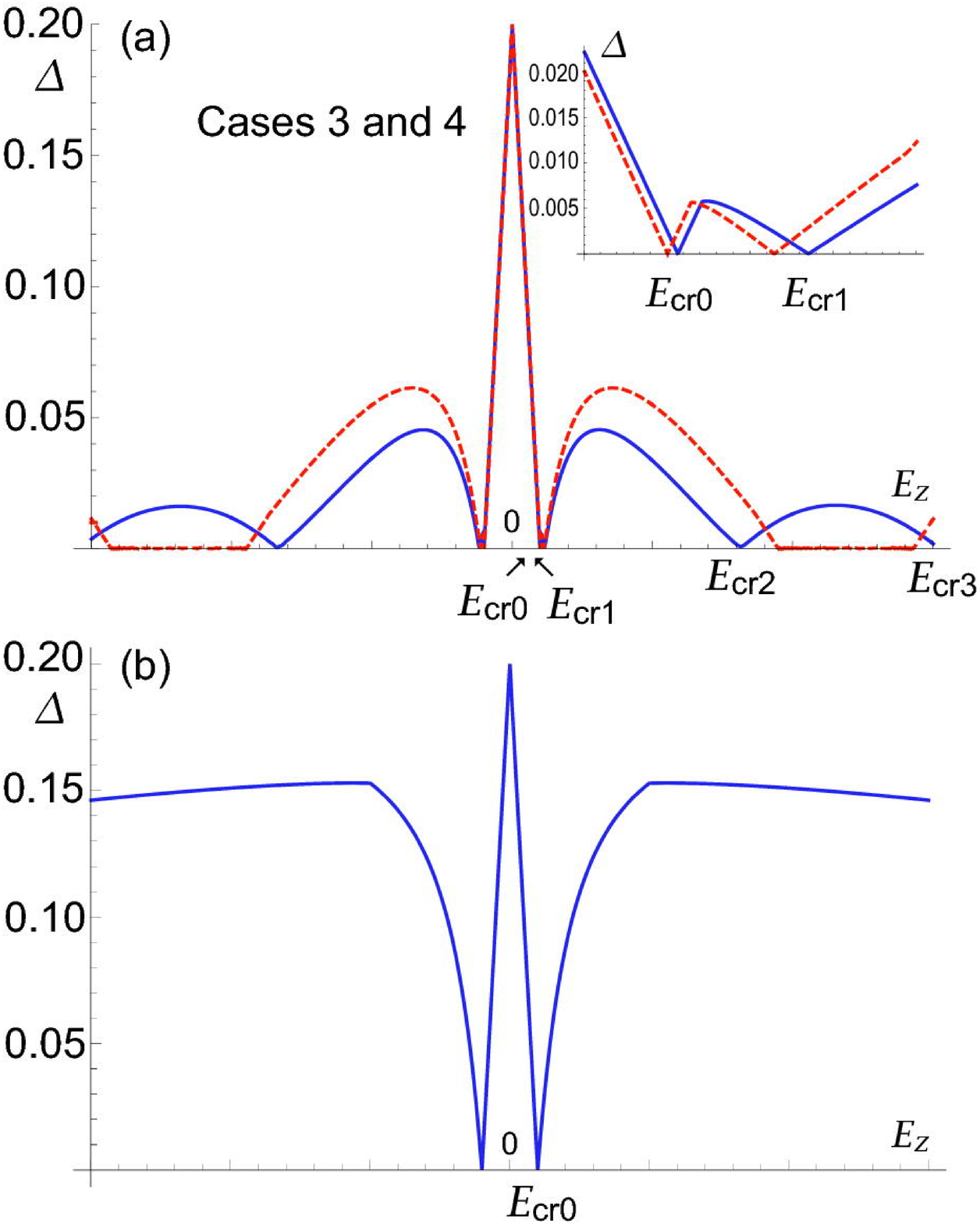}}
\caption{(Color online) (a) The band gap $\Delta $ of the backward stacking
silicene as a function of the electric field $E_{z}$ for the cases 3 and 4.
The gap closes at the critical points $E_{\text{cr}0},E_{\text{cr}1},E_{%
\text{cr}2}$ and $E_{\text{cr}3}$. Bilayer silicene is an insulator except
for these points. We note that the critical points $E_{\text{cr}2}$ and $E_{%
\text{cr}3}$ are missed in the effective two-band theory. (b) The band gap
in the limit $t_{3}\rightarrow 0$, where all the critical points collap to
one. Solid (dashed) curve is the band gap obtained based on the Dirac
Hamiltonian (\protect\ref{DiracHamilE}) by taking $\protect\lambda _{\text{R}%
}/\protect\lambda _{\text{SO}}=1/2$, $\protect\lambda _{\text{inter}}/%
\protect\lambda _{\text{SO}}=1/4$ and $\protect\lambda _{\text{SO}}/t=1/10$
for illustration to emphasize the effects ($\protect\lambda =\protect\lambda %
_{\text{inter}}=0$ for comparison). The band gap is identical for the cases
3 and 4.}
\label{FigBackEGap}
\end{figure}

\begin{figure}[t]
\centerline{\includegraphics[width=0.5\textwidth]{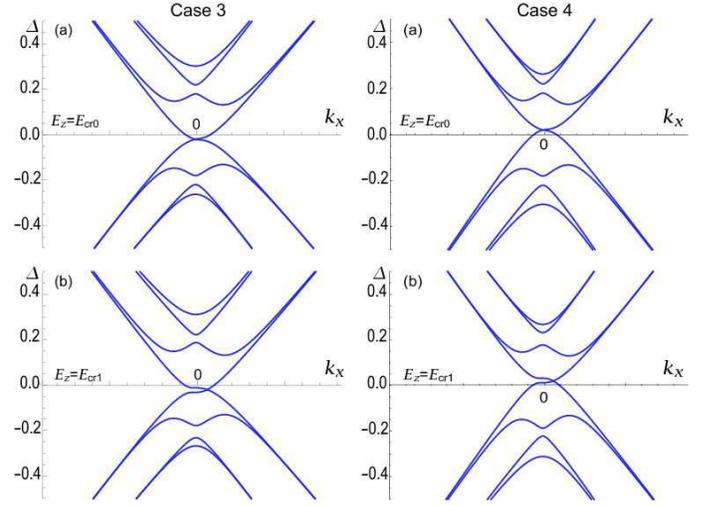}}
\caption{(Color online) Band structure of the backward Bernal (AB) stacking
bilayer silicene at the critical fields $E_{\text{cr}0}$ and $E_{\text{cr}1}$
for the cases (3) and (4). The gap structure $\Delta$ for the case 4 is
given by $-\Delta$ for the case 3. The gap closes at $k_{x}=0$ for $E_{\text{%
cr}0}$, but at $k_{x}\neq 0$ for $E_{\text{cr}1}$ due to the trigonal
warping. The Fermi level is the one at $E_{z}=0$.}
\label{FigBEc}
\end{figure}

The cases 3 and 4 are identical by reversing the whole system. In the case
of the backward Bernal stacking, the effective two-band Hamiltonian under
homogeneous electric field is derived as 
\begin{align}
H_{\text{eff}}^{\eta }=& h_{B}+h_{w}+m_{\text{D}}\left( 
\begin{array}{cc}
1 & 0 \\ 
0 & -1%
\end{array}%
\right) -\ell E_{z}\left( 
\begin{array}{cc}
1 & 0 \\ 
0 & 1%
\end{array}%
\right)   \notag \\
& -\frac{\hbar ^{2}v_{\text{F}}^{2}}{t_{\perp }^{2}}\left( LE_{z}+s\lambda _{%
\text{SO}}\right) \left( 
\begin{array}{cc}
k_{-}k_{+} & 0 \\ 
0 & -k_{+}k_{-}%
\end{array}%
\right)   \notag \\
& +\frac{\hbar ^{2}v_{\text{F}}^{2}}{t_{\perp }^{2}}\ell E_{z}\left( 
\begin{array}{cc}
k_{-}k_{+} & 0 \\ 
0 & k_{+}k_{-}%
\end{array}%
\right) ,  \label{HamilBack}
\end{align}%
with the Dirac mass%
\begin{equation}
m_{\text{D}}=LE_{z}-s\lambda _{\text{SO}}.
\end{equation}%
The energy spectrum has no electron-hole symmetry. The Fermi energy is moved
to $-\ell E_{z}$ by the presence of the electric field. The spectrum becomes
gapless at $E_{z}=E_{\text{cr}0}$ with%
\begin{equation}
E_{\text{cr}0}=\frac{s\lambda _{\text{SO}}}{L},  \label{CritiEb0}
\end{equation}%
which occurs at $k_{\pm }=0$. We can show that the
band gap is given as 
\begin{equation}
\Delta =2\lambda _{\text{SO}}-2L|E_{z}|,
\end{equation}%
for $|E|<E_{\text{cr0}}$. See also Fig.\ref{FigBackEGap} we derive by numerical
calculations.. We note that the energy spectrum is identical between the
cases 3 and 4. We show the band structure at the critical electric field in
Fig.\ref{FigBEc}(a). The band structure at the Fermi energy is
parabolic, which is reminiscence of bilayer graphene.

There exists one more critical field implied by the $2\times 2$ Hamiltonian,
where the spectrum become gapless. The condition is that the Hamiltonian (%
\ref{HamilForward}) itself becomes zero, as implies 
\begin{align}
& LE_{z}-s\lambda _{\text{SO}}-\frac{\hbar ^{2}v_{\text{F}}^{2}}{t_{\perp
}^{2}}\left( LE_{z}+s\lambda _{\text{SO}}\right) k_{-}k_{+}=0, \\
& \hbar v_{3}k_{+}-\frac{\hbar ^{2}v_{\text{F}}^{2}}{t_{\perp }}k_{-}^{2}=0.
\end{align}%
There are two solutions with $k_{y}=0$. The first one is $k_{x}=0$, where $%
E_{z}$ is given by (\ref{CritiEb0}). The new solution is given by solving 
\begin{equation}
k_{x}=\pm \frac{t_{\perp }\sqrt{LE_{z}-s\lambda _{\text{SO}}}}{\hbar v_{%
\text{F}}\sqrt{LE_{z}+s\lambda _{\text{SO}}}}=\frac{v_{3}t_{\perp }}{\hbar
v_{\text{F}}^{2}},
\end{equation}%
which yields 
\begin{equation}
E_{\text{cr}1}=\frac{(1+v_{3}^{2}/v_{\text{F}}^{2})s\lambda _{\text{SO}}}{%
(1-v_{3}^{2}/v_{\text{F}}^{2})L}.
\end{equation}%
There are two more solutions by performing $\pm 2\pi /3$ rotations in the ($%
k_{x}$,$k_{y}$) plane around ($0,0$) with the same critical field $E_{\text{%
cr}1}$ due to the trigonal symmetry.

Actually there are two more critical points, $E_{\text{cr}2}$ and $E_{\text{%
cr}3}$, which are present in the original Hamiltonian (\ref{HamilExpli})
with (\ref{DiagoEback}) but missed in the reduced two-band model (\ref%
{HamilBack}), as we demonstrate in Fig.\ref{FigBackEGap}. It is intriguing
that the band is flat for $E_{z}>E_{\text{cr}2}$. Note that $E_{\text{cr}%
1}\rightarrow E_{\text{cr}0}$ but $E_{\text{cr}2}\rightarrow \infty $ as $%
t_{3}\rightarrow 0$. Namely, these two critical points are generated by the $%
t_{3}$ hopping effect.

\subsection{Forward Bernal Stacking in Electric Field: Case 2, $AB$-$\tilde{A%
}\tilde{B}$ type}

\begin{figure}[t]
\centerline{\includegraphics[width=0.4\textwidth]{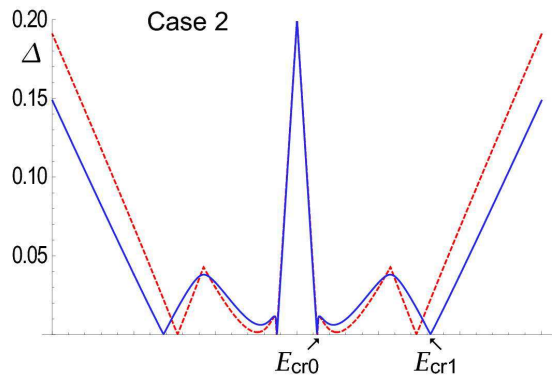}}
\caption{(Color online) (a) The band gap $\Delta $ of the forward stacking
bilayer silicene as a function of the electric field $E_{z}$ for the case 2.
The gap closes at the critical points $E_{\text{cr}0}$ and $E_{\text{cr}1}$.
Bilayer silicene is an insulator except for these points.}
\label{FigBABA}
\end{figure}

Finally we investigate the case 2. In this case, since we have $%
t_{3}>t_{\perp }$, the two-band Hamiltonian of the type (\ref{perturb}) is
unavailable. We are unable to present analytic formulas. We have calculated
the band gap numerically, which we show in Fig.\ref{FigBABA}.The number of
the critical electric field is found to be reduced to two. 

\section{Discussions}

Bilayer silicene has such a peculiar feature that its physical properties
are more akin to those of a topological insulator than those of a band
insulator though it has a trivial $\mathbb{Z}_{2}$ topological charge.
Namely it is characterized by a full insulating gap in the bulk and almost
helical gapless edges. It exhibits a similar behavior as monolayer silicene
under electric field $E_{z}$. Physical properties are very similar to those
of a topological insulator for $|E_{z}|<E_{\text{cr}0}$, while they become
those of a band insulator for $|E_{z}|>E_{\text{cr}0}$ with a transition
point $E_{\text{cr}0}$. The gap closes precisely at the transition point,
where bilayer silicene is a metal while monolayer silicene is a semimetal.

As we have discussed in the case of monolayer silicene\cite{EzawaNJP},
therefore, by applying an inhomogeneous electric field, we would be able to
create metallic regions anywhere within a bilayer silicene sheet where
helical zero-energy modes are confined. Furthermore, by rolling up a sheet,
we may consider a double-wall silicon-nanotube, and by applying an electric
field perpendicular to the tube axis, we would be able to create metallic
channels parallel to the tube axis where spin currents are conveyed\cite%
{EzawaSiTube}.

I am very much grateful to N. Nagaosa for many fruitful discussions on the
subject. This work was supported in part by Grants-in-Aid for Scientific
Research from the Ministry of Education, Science, Sports and Culture No.
22740196.

\end{document}